\documentclass{icrc2009}

\usepackage{graphicx}   
\usepackage[font=footnotesize]{subfig} 
\usepackage{fixltx2e}
\usepackage{url}

\newcommand{\shorttitle}[1]%
{\markboth{Proceedings of the 31\MakeLowercase{$^{st}$} ICRC, {\L}\'{o}d\'{z} 2009}{#1} }
\newcommand{\etal}{\MakeLowercase{\textit{et al. }}} 

\begin{document}
\title{Observation of shadowing of the cosmic electrons and positrons by the Moon with IACT}

\author{\IEEEauthorblockN{P. Colin\IEEEauthorrefmark{1},
			  D. Borla Tridon\IEEEauthorrefmark{1},
			  D. Britzger\IEEEauthorrefmark{1},
			  E. Lorenz\IEEEauthorrefmark{1},
              R. Mirzoyan\IEEEauthorrefmark{1},
			  T. Schweizer\IEEEauthorrefmark{1},
			  M. Teshima\IEEEauthorrefmark{1}\\
              for the MAGIC collaboration}
                            \\
\IEEEauthorblockA{\IEEEauthorrefmark{1} Max-Planck-Institut f\"ur Physik, F\"ohringer Ring 6, 80805 Munich, Germany}
colin@mppmu.mpg.de}

\shorttitle{Colin \etal electron Moon Shadow observation}
\maketitle

\begin{abstract}
Recent measurements of the cosmic-ray electron ($e^-$) and positron ($e^+$) fluxes
show apparent excesses compared to the spectra expected by standard
cosmic-ray (CR) propagation models in our galaxy.
These excesses may be related to particle acceleration in local
astrophysical objects, or to dark matter annihilation/decay.
The $e^+$/$e^-$ ratio (measured up to $\sim$100\,GeV) increases unexpectedly above 10\,GeV
and this may be connected to the excess in all-electron ($e^+$+$e^-$)
flux at 300-800\,GeV. Measurement of this ratio at higher energies is a key parameter
to understand the origin of these spectral anomalies.
Imaging Atmospheric Cherenkov Telescopes (IACT) detect electromagnetic air showers above 100\,GeV, but with this technique, the discrimination between primary $e^-$, $e^+$ and diffuse $\gamma$\,rays is almost impossible. However, the Moon and the geomagnetic field provide an incredible opportunity to separate these 3 components. Indeed, the Moon produces a 0.5$^\circ$-diameter hole in the isotropic CR flux, which is shifted by the Earth magnetosphere depending on
the momentum and charge of the particles. Below few TeV, the $e^+$ and $e^-$ shadows are shifted at $>$0.5$^\circ$ each side of the Moon and the $e^+$, $e^-$ and $\gamma$-ray shadows are spatially separated. IACT can observe the $e^+$ and $e^-$ shadows without direct moonlight in its field of view, but the scattered moonlight induces a very high background level.
Operating at the highest altitude (2200\,m), with the largest telescopes (17\,m) of the current IACT, MAGIC is the best candidate to reach a low energy threshold in these peculiar conditions. Here we discuss the feasibility of such observations.
\end{abstract}

\begin{IEEEkeywords}
 cosmic ray, electron, positron, Cherenkov telescope
\end{IEEEkeywords}

\section{Introduction}
The interaction of cosmic rays (CR) with the interstellar medium induces
secondary particles which can be detected on Earth.
Models of CR propagation in the galaxy can explain the measured relative abundance of different components ($\bar p$/$p$, B/C) as well as the diffuse $\gamma$-ray emission at the GeV regime \cite{Strong04}.
Generally, these models predict a smooth all-electron spectrum decreasing above 30\,GeV faster than a power law $E^{-3}$ and a $e^+$/$e^-$ ratio decreasing slowly with the energy. However, the recent measurements of the cosmic electrons and positrons show a different picture
(see Fig~1).
PAMELA \cite{Adriani09} reported an increasing $e^+$/$e^-$ ratio above 10\,GeV, in agreement with the previous results of HEAT \cite{Beatty04} and AMS-01 \cite{Aguilar07}.
Between 30\,GeV and 500\,GeV, Fermi \cite{Abdo09} measured an all-electron spectrum following a power-law with a spectral index of -3.0 and ATIC \cite{Chang08} reported an even harder spectrum with bump between 300\,GeV and 700\,GeV. At higher energy, both ATIC and HESS \cite{Aharonian08} measured a break in the spectrum around 800\,GeV.
There is a small discrepancy between experiments, but they all measured a spectrum
that cannot be explained easily by conventional CR-diffusion models.

 \begin{figure}[!t]
  \centering
  \includegraphics[width=6.5cm]{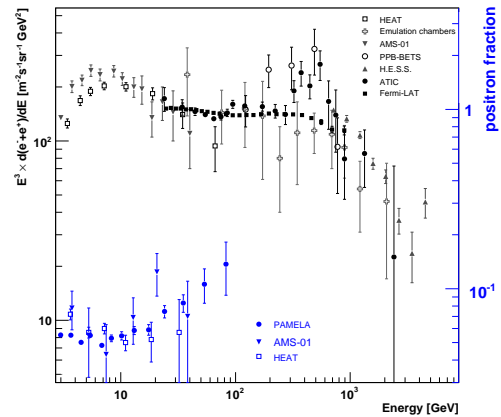}
  \caption{Recent measurements of the cosmic electrons and positrons. See references in the text.}
  \label{fig1}
 \end{figure}

The anomalies in the electron and positron fluxes are generally interpreted as a new component with a harder spectra and a higher $e^+$/$e^-$ ratio than the flux induced by the CR propagation and interaction in the galaxy. Many scenarios involving dark matter (annihilation/decay), pulsars or special CR models have been proposed to interpret the data. The $e^+$/$e^-$ ratio predicted
above 100\,GeV by these models can be very different. Measurement of this ratio at higher energies (up to the electron spectrum break and even over) is essential to establish a connection between the PAMELA excess and the ATIC bump and to discriminate between the different models.
With more statistic, PAMELA may measure the $e^+$/$e^-$ ratio up to 300\,GeV but could probably not reach higher energy. Few years after its launch (foreseen in 2010), AMS-02 \cite{AMS02} may measure the $e^+$ flux up to 1\,TeV in the most optimist case.
Using the Moon shadowing effect, IACT observatories could measure or constrain the $e^+$/$e^-$ ratio around 1\,TeV. With the MAGIC telescopes, the energy range between 300\,GeV and 1\,TeV should be observable. This range could be even larger with the next generation of IACT.

\section{The Earth/Moon spectrometer}

The Earth/Moon system forms a spectrometer where the Earth magnetosphere deflects the trajectory of any coming particle depending on its charge and momentum (equivalent to its energy for an ultra-relativist particle) and the Moon creates a hole in the isotropic CR flux corresponding to the CRs which would go through the Moon to reach an observer on Earth.
The missing flux scales as the solid angle of the Moon $(6.6\pm0.8)\times 10^{-5}$\,sr ($\sim$0.5$^\circ$ diameter) and varies of $\pm12\%$ as a function of the observer-Moon distance.
For example, the proton flux above 1\,TeV being about $6\times10^{-2}$m$^{-2}$s$^{-1}$sr$^{-1}$, the mean proton flux blocked by the Moon is about $4\times10^{-6}$m$^{-2}$s$^{-1}$. This corresponds to $\sim$20 times the Crab nebula $\gamma$-ray flux above 1\,TeV. For electrons, the missing flux is given Table~1 for several energies and for different spectrum hypothesis (the missing flux is shared between $e^-$ and $e^+$).

 \begin{table}[!h]
  \caption{Mean missing flux of all-electron due to the Moon shadowing in Crab nebula $\gamma$-ray-flux unit.}
  \label{table_simple}
  \centering
  \begin{tabular}{|c|c|c|c|c|}
  \hline
   Flux\footnotemark & $>$300\,GeV & $>$500\,GeV&  $>$700\,GeV&  $>$1\,TeV \\
   \hline
    ATIC bump & 5.8\% & 3.5\% & 2.0\% & 1.1\% \\
    Fermi/HESS & 4.2\% & 3.0\% & 2.0\% & 1.1\% \\
    CR model & 2.6\% & 1.7\% & 1.3\% & 1.0\% \\
  \hline
  \end{tabular}
  \end{table}
  \footnotetext{Crab flux: $dN/dE=3\times10^{-7}(E/TeV)^{-2.5}$
  m$^{-2}$s$^{-1}$TeV$^{-1}$}

The position of the CR-flux hole (Moon shadow) depends on the charge and energy of the particles. For neutral CRs (like diffuse $\gamma$\,rays), it lies at the actual Moon position. For charged CRs, the Moon shadow is shifted perpendicularly to the geomagnetic field along an axis close to an East-West orientation. Negative and positive CRs are shifted respectively eastward and westward. As most of the CRs are positive particles (atom nuclei), the all-CR Moon shadow is asymmetric with a larger deficit at the west side of the Moon.
The Moon (and Sun) shadowing effect is used by ground-based EAS detector
to estimate their angular resolution. Experiments with the lowest energy threshold,
such as L3+C \cite{l3c2004} and Tibet-III \cite{tibet2007},
have detected the East-West asymmetry of the Moon shadow. They deduced the best upper limits
($\sim$7\%) on the $\bar p$/$p$ ratio at the TeV regime from these measurements.

 \begin{figure}[!t]
  \centering
  \includegraphics[width=6.5cm]{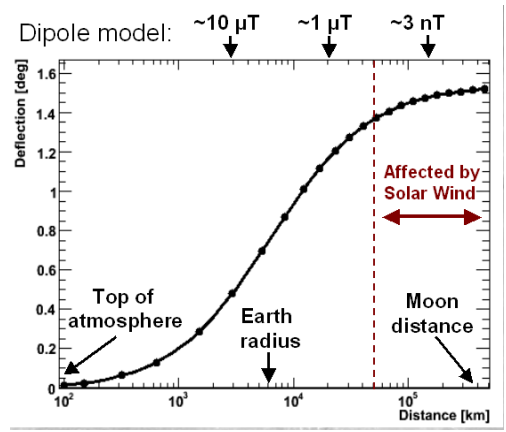}
  \caption{Deviation angle of a 1\,TeV-proton trajectory in the centred-dipole model as a function of the covered distance from a point on Earth. The geomagnetic field intensity of this model at several distances is indicated at the top.}
  \label{fig2}
 \end{figure}

The amplitude and direction of the apparent deviation of the Moon shadow
depends on the magnetic field between the Moon and the observer.
The geomagnetic field is intensively studied by the international association of geomagnetism and aeronomy (IAGA) which releases a reference model (IGRF) every 5 years \cite{IGRF}. The IGRF provides models for the past and the future (next 5 years) taking into account the long-term variation of the main field. However, it does not account for short-time-scale variations like a magnetic storm. More complete models including the magnetosphere deformation induced by the solar wind, have been developed to study trajectories of low energy CRs in the Earth vicinity \cite{smart2000}.
At the first order, the geomagnetic field can be considered as a magnetic dipole at the centre of the Earth. The difference between the centred-dipole model and the real geomagnetic field varies between $\pm10\%$. Thus, the centred-dipole model is good enough to draw conclusions on the feasibility. Figure~2 shows the deviation angle of a 1\,TeV-proton trajectory
in this model as a function of the covered distance from a point on the Earth surface. Main part of the deviation happens between the top of the earth atmosphere and the distance of the Moon. Thus, the Moon shadow position is not affected by the uncertainty on the altitude of the first interaction in the atmosphere. However, a small part (not negligible) of the deviation may be affected by the interplanetary magnetic field (2-5\,nT at the Earth distance).
This effect cannot be easily predicted because it depends on the solar activity, but it can be estimated afterward from the solar wind measurements. This correction should be less than 10$\%$.

The amplitude and direction of the deviation depend on the longitude and latitude of the observer as well as on the azimuth and zenithal angles of the Moon (for the observer). At the latitude of MAGIC, the deviation of particle with a charge $Z$ and an energy $E$ varies about from $Z$/$E\times$1.3$^\circ$ at zenith, up to $Z$/$E\times$3$^\circ$ at 80$^\circ$ from zenith in the East-West direction (with $Z$ in elementary charge unit and $E$ in TeV). The positions of the $e^-$ and $e^+$ shadows for the energy range of the all-electron-spectrum feature (300\,GeV-1\,TeV) are few degrees from the Moon centre. Figure~3 shows the shadow positions in the sky for a typical observation toward the East at 45$^\circ$ from zenith.

 \begin{figure}[!t]
  \centering
  \includegraphics[width=6.5cm]{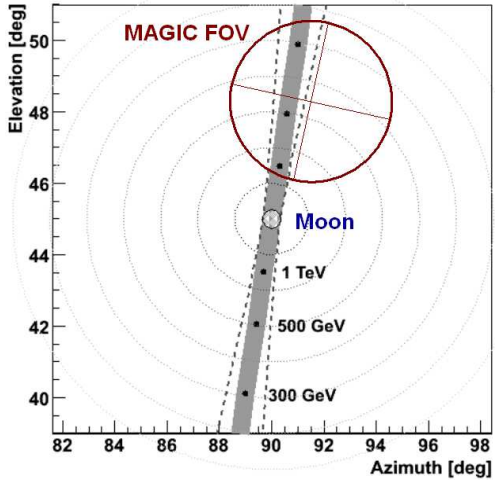}
  \caption{Positions of the Moon shadows for an observer at the MAGIC telescopes site with a rising Moon at 45$^\circ$ of elevation. The $e^-$ shadow is below the Moon (eastward) and the $e^+$/proton shadow is above the Moon (westward). The dashed lines represent the position uncertainty induced by a 10$\%$ error on the geomagnetic field. The light-grey dotted circles are the curves of iso-distance to the Moon with 1$^\circ$ step. The large target shows the possible position of the MAGIC field of view during $e^+$/proton shadow observation at the energy range 300\,GeV-1\,TeV.}
  \label{fig3}
 \end{figure}

\section{Observation with IACT}

In the 90s, the ARTEMIS experiment \cite{artemis2001} tried to measure the Moon-shadow effect with a Cherenkov telescope in order to measure the $\bar p$/$p$ ratio.
They equipped the Whipple-10m telescope with a special 109-pixel camera exclusively sensitive to UV light between 200-300\,nm. At this wavelength range, the moonlight is absorbed by the ozone layer and does not induce background signal.
The analysis was based on a simple rate counting hinged on the position of the shower
centroid relatively to the Moon position. The angular resolution achieved ($\geq1^\circ$) was
much larger than what is achieved by the recent IACT ($\geq0.1^\circ$) using image-shape analysis and stereoscopic reconstruction technique.
The expected deficit in the ARTEMIS data due to the proton shadow has been estimated by Monte Carlo simulation at only $1\%$ of the background rate. Such level of background control is almost impossible and the experiment did not detect any significant deficit in the CR flux.

Since ARTEMIS, no other tentative was done to measure the Moon shadow with IACT in spite of the big progress of this technique during the last decade. Actually, the main problem of this observation is the dazzling background light induced by the scattered moonlight.
The MAGIC telescopes use low gain PMTs with only 6 dynodes, that can be operated at high light background levels without any damage. MAGIC-I observes under moonlight for long time. The
recent upgrade of the data-acquisition system (2\,GSample/s) associated to the time-image-cleaning technique makes the MAGIC-I performance very robust. Recent study shows that the sensitivity above 200\,GeV is almost not affected by an increase of the night sky background 5 times higher than the normal dark-sky level \cite{Britzger09}.
The intensity of the background depends on the Moon phase, the distance to the Moon and the elevation. The Moon-shadow observation would be dramatically closer to the Moon than the standard MAGIC observations under moonlight. Some test made with MAGIC-I shows that the night sky background
at 3.5$^\circ$ from the Moon is about 40 times higher than the dark-sky level for a half-Moon at 45$^\circ$ of elevation. Due to the security limit to operate MAGIC (not yet precisely established), Moon-shadow observations are probably not possible with a Moon phase larger than 50$\%$ (the phase is defined as the lit-area fraction of the lunar disc). Some hardware modifications, such an UV-filter in front of the camera, may change this limit and increase the signal to noise ratio.

The energy threshold of IACT rises strongly with the distance to zenith. In order to keep an energy threshold below 300\,GeV with MAGIC, the zenith angle must be less than $\sim$50$^\circ$ (the exact limit depends also on the Moon phase).
As the small Moon phase corresponds to low angle between the Moon and Sun, a small-phase Moon is rarely at high elevation during the night. There are only about 30\,h/year with
a $<$50\%-phase Moon at less than 50$^\circ$ from zenith during astronomical nights (Sun $>$18$^\circ$ below the horizon). When this happens, the Moon is toward East (rising just before the Sun) or toward West (falling just after the Sun set). As it is shown Figure~3, in this configuration the Moon shadow spreads vertically above and below the Moon. The shadow above the moon is in better position to be observed because it is closer to zenith (lower energy threshold) and the bright side of the Moon is its bottom part (less background light).
Thus, the best period to observe the $e^-$ Moon shadow is the begin of the night around the spring equinox and for the $e^+$/proton shadow, it is the end of the night around the fall equinox.



The deviation angle depends on the energy, so the spatial shape of the Moon shadow for a given reconstructed energy depends on the energy resolution.
The energy resolution of the MAGIC telescopes in stereoscopic mode \cite{Colin09} is about 15\%
for $\gamma$ rays above 200\,GeV. Performance for electrons is similar. Thus the Moon shadow is elongated along the deviation axis by $\sim$15\% of the deviation angle. As the typical deviation angle is 3-4$^\circ$ at 500\,GeV, the Moon shadow at this energy is spread out by $\sim$0.5$^\circ$ and its size grows twice of the actual lunar disc. The shadow is more extended at lower energies and more compact at higher energies. The angular resolution is about half of the Moon radius, so the dispersion in perpendicular direction to the deviation axis is much smaller.
The Moon shadow is then strongly extended in only one direction and it can be contained in one half of the camera. The other half could be used to estimate the CR-background rate in a very similar elevation and background light condition. Observation in Wobble mode with a 0.6$^\circ$ offset (as shown Figure~3) seems a good strategy.

Electrons are the main negative component of the CR flux and the deficit induced by the Moon shadow should be easily identifiable. The anti-hadron flux at the TeV energy is unknown but the expected flux is far bellow the best upper limit on $\bar p$/$p$ (7\%). This upper limit could be even improved by the Moon shadow observation with IACT.
Moreover anti-hadrons would be strongly suppressed by electron selection cuts (similar to the $\gamma$-ray selection \cite{Colin09}), and the few anti-proton showers that could mimics electromagnetic showers, are generally more energetic than the reconstructed energy and they would be less deflected by the geomagnetic field.
For positrons, instead, the discrimination with hadron Moon shadows is more difficult.
Some protons transferring most of the energy to a $\pi^0$ at the first interaction may produce flux deficit very similar to $e^+$ shadow. For helium nuclei, the ratio of real energy to the reconstructed one is similar to their charge of 2. So, the assumed deflection could be close to the real one. Hadrons with $Z$$>$2, however, should not be a problem because they are strongly rejected and much more affected by the magnetic field.
The $e^+$ flux measurement will rely on the detailed proton and helium Monte Carlo simulations. These simulations could be confronted to hadron Moon shadow analysis using reconstruction and selection cuts adapted to proton or helium.

The performance of the MAGIC telescopes for extended sources under bright moonlight is still under study. We estimate that the $e^{-}$ Moon-shadow detection should requires between 10\,h and 50\,h of observations. As only $\sim$15\,h/year are available (30\,h/year shared between $e^-$ and $e^+$), we should need between 1 and 4 years to measure the $e^{-}$ spectrum using the Moon shadowing effect.
Concerning the $e^{+}$ shadow detection, more simulations are needed to draw any conclusion.
On the other hand, the $e^{-}$+$e^{+}$+diffuse-$\gamma$-ray spectrum can be estimated thanks to the very good $\gamma$-electron/hadron separation of IACT at high energy \cite{Aharonian08}. Then, upper limits on the $e^{+}$ and diffuse-$\gamma$-ray fluxes could be deduced from the $e^{-}$ Moon shadow detection.

With the next generation of IACT, the sensitivity should strongly improve and the $e^{-}$ shadow detection time may decrease dramatically ($<$5\,h). The accessible energy
range should also widen at lower energies thanks to larger telescopes with larger field of views,
and at higher energies thanks to observations at large zenith angles (where the deviation angle is larger) with larger-effective-area array. An energy range from 150\,GeV to 3\,TeV seems possible.
Obviously, the hardware design must allow observation with very high light background. A data acquisition system with a fast sampling is also required to prevent the Cherenkov image degradation due to the high background light. The improvement of the signal to noise ratio with UV-filters should be also studied.
Observation of the electron and proton shadowing by the Moon could also provide a direct and independent energy calibration of IACT observatories and a good test for Monte Carlo simulation program.

\section{Conclusion}
Using the Moon shadow effect, the IACT arrays have a chance to measure anti-mater/mater ratios in the cosmic ray flux at energies hardly accessible with satellite experiments. The very good electron/proton separation performance of the IACT technique should allow us to measure or constrain both $e^+$/$e^-$ and $\bar p$/$p$. The $e^+$/$e^-$ ratio at the TeV regime is particulary interesting as features were reported in the all-electron spectrum and as a positron excess is measured between 10\,GeV and 100\,GeV.
For $e^{-}$, the expected missing flux (3-5\% Crab nebula above 300\,GeV) and the spatial extension ($\sim$\,0.5$^\circ$$\times$1$^\circ$) of the Moon shadow is in the reach of the current IACT.
For $e^{+}$, the detection will be more difficult as the missing flux should be lower and the hadron Moon shadows may produce similar deficit.
The high light background induced by the scattered moonlight makes this observation challenging. The IACT cameras must have been design to operate with a very high night-sky-background light without damage. With the MAGIC telescopes, such an observation seems feasible with enough sensitivity to detect the $e^{-}$ shadow in less than 50\,h. But because of the short time slot per year, this may take several years.

\section{Acknowledgements}
We thank the Instituto de Astrofisica de Canarias for
the excellent working conditions at the Observatorio del
Roque de los Muchachos in La Palma. The support of
the German BMBF and MPG, the Italian INFN, and
Spanish MCINN is gratefully acknowledged. This work
was also supported by ETH Research Grant TH 34/043,
by the Polish MNiSzW Grant N N203 390834, and by
the YIP of the Helmholtz Gemeinschaft.

\end{document}